\documentstyle[12pt,aasms4]{article}
\newcommand{\etal}{{\it et al.} }
\newcommand{\asca}{{\it ASCA} }

\newcommand{\exosat}{{\it EXOSAT} }

\newcommand{\mcg}{MCG $-$6$-$30$-$15 }
\newcommand{\epft}{$EPF(T)$ }

\begin{document}

\title{A SIMPLE NEW METHOD FOR ANALYSING GAPPED TIME-SERIES:
SEARCH FOR A HIGH-FREQUENCY CUT-OFF IN THE X-RAY
POWER SPECTRUM OF THE SEYFERT GALAXY MCG $-$6$-$30$-$15}
 
\author{T. Yaqoob\altaffilmark{1,2},
B. McKernan\altaffilmark{3},
A. Ptak\altaffilmark{4},
K. Nandra\altaffilmark{1,5},
P. J. Serlemitsos\altaffilmark{1}}

\vspace{3cm}

\altaffiltext{1}{NASA/ Goddard Space Flight Center, Laboratory for High
Energy Astrophysics, Greenbelt, MD 20771, USA.}
\altaffiltext{2}{With the Universities Space Research Association.}
\altaffiltext{3}{Department of Physics, University of Leeds, Leeds, UK.}
\altaffiltext{4}{University of Maryland, College Park, MD 20742.}
\altaffiltext{5}{NAS/NRC Research Associate.}

\begin{abstract}

We present a new, simple method for analysing gapped time-series which
is particularly suited to probe the
X-ray power spectra of active galactic nuclei (AGN).
The method is applied to a four-day observation
of the Seyfert 1 galaxy MCG $-$6$-$30$-$15
with \asca ({\it Advanced Satellite for Cosmology and Astrophysics}).
MCG $-$6$-$30$-$15 is well-known
for rapid, large amplitude X-ray variability and was one of the
first AGN to show evidence of relativistic effects in its Fe-K line
profile. In this source, our method probes the power spectrum in the
range $10^{-3}$ Hz to $5 \times 10^{-2}$ Hz, thus
extending the high-frequency coverage by a factor of over 50 compared to
previous studies. The \asca data rule out a cessation of variability
up to $1.6 \times 10^{-2}$ Hz at greater than 90\% confidence. 
The power-law index of the power spectrum in this range is
consistent with $\sim -1.4$. 
At $2.1 \times 10^{-2}$ Hz
we
can place an upper limit of 7\% on the intrinsic fluctuations 
of the source luminosity at greater than 99\% confidence. 
Above this frequency Poisson noise becomes dominant but we can 
place an upper limit of 10\% on intrinsic fluctuations up 
to $5 \times 10^{-2}$ Hz at 98.9\% confidence.
The data argue against orbital modulation as the origin of the
variability since it would imply upper limits on the mass of the
putative black hole of less than $5 \times 10^{5} M_{\odot}$,
and therefore super-Eddington
luminosities. 
\end{abstract}

\keywords{methods: data analysis -- galaxies:active - galaxies:
individual: MCG $-$6$-$30$-$15 -- X-rays: galaxies}

\section{Introduction}
\label{intro}

The search for characteristic timescales in
the X-ray variability of active galactic nuclei (AGN) has
been a long-standing problem.
The X-ray power spectra 
have been difficult to measure due to a  lack of the combination 
of instrument sensitivity and long, uninterrupted
exposures. So far, \exosat has provided the best measurements
(see Lawrence \& Papadakis 1993; Green, McHardy and Lehto 1993;
Czerny and Lehto 1997, and references
therein).  
Green \etal (1993), from an analysis of the entire \exosat AGN
database found only nine sources with data of sufficient
quality to measure the power spectrum, six of these in the range
$10^{-4.5}-10^{-3.5}$ Hz and the remaining three 
(NGC 4051, NGC 5548 and MCG $-$6$-$30$-$15) in the range
$10^{-5}-10^{-3}$ Hz. Green \etal (1993) and Lawrence
and Papadakis (1993) found the power spectra to be featureless
with mean power-law indices $-1.72 \pm 0.52$ and $-1.55 \pm 0.09$
respectively. No breaks in the power spectra were found in
the measured frequency ranges. However there have been 
reports of peaks in the \exosat X-ray power spectra of three AGN;
at $1.7 \times 10^{-4}$ Hz and/or $7.9 \times 10^{-5}$ Hz
in NGC 4151 (Fiore, Massaro, and Perola, 1989);
at $\sim 2 \times 10^{-3}$ Hz in NGC 5548 (Papadakis and Lawrence, 1993), 
and at $\sim 4 \times 10^{-4}$ Hz in NGC 4051  
( Papadakis and Lawrence, 1995; see also Bao and 
Ostgaard 1994). 
All of these have yet to be confirmed with other missions. 
The problem is that 
after the demise of \exosat 
and its highly eccentric orbit, the orbits
of subsequent satellites have not been able to produce 
such long, essentially uninterrupted 
data streams. Although sophisticated methods have been devised 
to analyse time-series with gaps (e.g. see Merrifield and McHardy,
1994 and references therein), interpretation of the resulting
power-spectra, considering the complexities
introduced by the `window-function', is not straight-forward. 

The only other 
reports of `characteristic timescales' have come from 
a long-term variability analysis of NGC 4151 which showed evidence,
albeit model-dependent, of three timescales, at
$\sim 5,300- 53,000$ s, 14 days and 183 days (see 
Papadakis and McHardy, 1995).
So far, a high-frequency cut-off in the X-ray power spectra
of AGN has not been observed. A search for this cut-off is important
because it places hard physical constraints on the X-ray emission 
mechanism and/or the size of the region, and ultimately  the
mass of the putative black hole. So far, the best estimate
of the minimum X-ray variability timescale has come,
for only a few bright AGN, from
the points where their \exosat power spectra  
hit the Poisson noise (Green \etal 1993). These timescales 
were all greater than $10^{3}$ s.
In this paper we extend the search for a 
high-frequency cut-off in the power spectrum down to timescales
of 20 s ($5 \times 10^{-2}$ Hz) in the Seyfert 1 galaxy MCG $-$6$-$30$-$15,
utilizing high-sensitivity data taken with {\it ASCA}, 
using a new method for analysing 
gapped time-series.

\section{The Excess Pair Fraction}
\label{epfsec}

We define the `Excess Pair Fraction' 
( hereafter $EPF$) as a function of the bin size, $T$, of a given
lightcurve. For each bin size we count the total number of adjacent
pairs of {\it non-empty} bins and then compute the fraction of those
pairs in which the number of counts in the
two bins differ by, or more than, the sum of
the square roots of the counts.
This fraction is $EPF(T)$. 
In other words, if we observe
$C_{1}$ counts in the first bin of a pair, for the second bin to
contribute to \epft it must have $C_{2}$ counts where
$|C_{1} - C_{2}| \ge \sqrt{C_{1}} + \sqrt{C_{2}}$. This has the
solution that $C_{2}$ must {\it not} lie inside the range
$1+ C_{1} \pm 2 \sqrt{C_{1}}$. The second bin is used
as the first bin of the next pair tested, and so on.
This definition of $EPF$ is equivalent to saying that in the Gaussian
limit, the counts in each bin of a pair must be separated by at least
the sum of their standard deviations.

By definition, the $EPF$ characterizes the
source variability on the timescale of the bin size and nothing
in the definition requires uninterrupted data or inventing
fictitious data in the gaps. All that is
required is that there be enough adjacent pairs of bins.
It is different to a $\chi^{2}$ test 
against a constant  hypothesis as a function of bin size because
the $\chi^{2}$ test is {\it global} in the sense that it searches
for a variability trend over the entire observation
relative to the mean level. On the other hand, the definition
of \epft does not refer to a mean,
but rather, {\it explicitly measures the bin-to-bin variability}. 
For example, if a source
showed significant variations on 10 s timescales but no longer term
trends,
$EPF(10)$ would be larger than $EPF(1000)$ because the
variability over $\sim 1000$ s would be `washed out'. The definition
of \epft is also manifestly different to the excess variance
(e.g. Nandra \etal 1997) which probes the {\it integrated} power spectrum.
 
Figure 1 shows the results of 
several simulations of the $EPF$ using $10^{6}$ bin-pairs for
each case. The counts for each bin are drawn from a Poisson distribution
in which
the mean count rate of the
source changes from $\mu_{0}$ to $(1+f \eta) \mu_{0} $
between two  adjacent bins in a pair, where $\eta$ is a
uniform random deviate between $-1$ and $+1$.
Curves of $EPF(\mu)$ are shown for various values of $f$ in the range
0.0--1.0 where $\mu \equiv \mu_{0} T$. 
For clarity, only the case of a constant source ($f=0$) is shown 
down to $\mu < 10$ (filled circles). As $\mu$ tends to zero, $EPF(\mu)$ for
all $f$ tends to zero because of the condition in the definition of
$EPF(\mu)$ which excludes bin-pairs containing one or more empty bins. 
For a constant source, $EPF(\mu)$ is almost constant when 
we are well into the Gaussian regime ($\mu \gg 10$).

Figure 2a shows  realistic
simulations of \epft for a $\sim 4.2$ day observation 
of a constant source with a count rate of 1.0 ct/s. The time-intervals
for data-accumulation were identical to the good-time-intervals (GTI)
obtained
from a real  \asca observation of the Seyfert 1
galaxy \mcg (see Figure 3 and \S \ref{appl}).
However, as is evident from the definition of  $EPF(T)$,
the results do not depend at all on the GTI
and 
$EPF(T)$ is consistent with the expectation
for a constant source observed with no gaps (solid line, taken from the
$f=0$ simulations in Figure 1).
As shown in Figure 1, $EPF(T)$ can be extremely sensitive to variability.
The 1.0 ct/s constant source in the above simulation was modulated
with a sine wave of only 3\% amplitude (6\% peak-to-peak) and a 
period of 367 s. The resulting \epft is shown in Figure 2b. Not
only is the variability easily detected, but the periodicity
makes characteristic imprints on $EPF(T)$. Namely, $EPF(T)$ has
minima at times equal to the period and its harmonics. In this example,
we would be able to determine that 367 s is the fundamental period 
because no minimum is seen at half of this period.

\section{Application to MCG $-$6$-$30$-$15}
\label{appl}

The Seyfert 1 galaxy \mcg ($z=0.008$) has been extensively studied at 
X-ray and other wavelengths. 
It is one of the first
AGN in which evidence for relativistic effects on the Fe-K line
profile was found (Tanaka \etal 1995).
\asca observed MCG $-$6$-$30$-$15 for over 4 days starting 1994, July 23  
(these data have already been discussed in other contexts by
Iwasawa \etal 1996, and Otani \etal 1996).      
MCG $-$6$-$30$-$15 is highly variable in the X-ray
band; the maximum 2--10 keV luminosity was over a factor of 7
greater than the minimum luminosity during the four-day
observation, in the range 
$3.8-28.2 \times 10^{42} \ \rm ergs \ s^{-1}$
($H_{0} = 50 \ \rm km \ Mpc^{-1} \ s^{-1}$ and $q_{0} = 0$). 
The 0.5--10 keV lightcurve obtained from 
the SIS0 instrument is shown in Figure 3, binned at 512 s
(see Tanaka \etal 1994 for details of the \asca instrumentation). 
The reader is referred
to Nandra \etal (1997)
for details of the \asca data reduction procedures which are very similar
to the ones adopted here.

We constructed lightcurves with a range of time-bin sizes
from $T=2$ s up to $T=1636$ s. All four instruments
aboard \asca were utilized. Partially-exposed bins
were rejected. Data in the 0.5--10 keV
band were used for the two SIS instruments and in the
0.7--10 keV band for the two GIS instruments. 
However, since the time resolution of SIS data is only
4 s, we did not use SIS data for $T < 32$ s, allowing sufficient
time for several CCD-readouts. The time resolution of the GIS data
is 2 s or better (depending on bit-rate).
$EPF(T)$ was then computed for each lightcurve.
Note that the lightcurves were {\it not} background-subtracted.
Even when the source has its minimum intensity (see Figure 3),
the background constitutes less than 11 \% of the total
counts and is not strongly variable (see Gendreau \etal 1995).

If we average \epft from different instruments
we must model \epft independently for the different instruments and
then average the models in the same way as the data. This is because  
in general the count rates and GTIs may be different for data from
the different instruments, and 
\epft is,
for a given $T$, 
fundamentally a function of the total number of counts per bin
($\mu$). 
Figure 4 shows, using the
entire \mcg observation,
\epft averaged over four instruments for $T \ge 32$ s
and over two instruments otherwise.
The error bars are obtained from the square root of the number
of bin-pairs contributing to \epft
divided by the total number of bin-pairs tested.
For a given mean count rate, $\mu_{0}$, we can convert the
$EPF(\mu)$ models in Figure 1 to \epft by interpolation,
using $\mu = \mu_{0}T$. The mean count rates were 1.83, 1.49, 0.96,
and 1.14 ct/s 
for SIS0, SIS1, GIS2 and GIS3 respectively.
In Figure 4,
the predicted \epft are shown for a source in which
the bin-to-bin ratio of mean count rates is independent of $T$ and uniformly
distributed in the range  $1 \pm f$, for $f = 0$ (constant source),
$f = 0.07$, and 0.20 (dotted curves). It can be seen that the measurement
errors cause the \epft to hit the Poisson noise below $T \sim 50$ s.
At $T=48 s$ ($2.1  \times 10^{-2}$ Hz)
we can say that the source variability is less than
7\% ($f<0.07$) at greater than $2.80 \sigma$, or 
greater than 99\% confidence. 
At 
$T = 20$ s, ($5 \times 10^{-2}$ Hz), approximately the smallest bin size
obeying Gaussian statistics, 
we can say that $f < 0.10$ at $2.53 \sigma$ (98.9 \% confidence).
At $T = 1000$ s ($10^{-3}$ Hz) the variability is $\sim 20\%$
($f \sim 0.20$). Note that the low-frequency  measurement limit for
\epft  
depends on the typical duration of an uninterrupted
data train (e.g. satellite orbit) 
and on the total duration of the observation.

Detailed fitting of the $EPF$ with models of power spectra
will be presented elsewhere. What is important here is that
the variability amplitude decreases 
steadily and smoothly down to at least 64 s ($1.6 \times 10^{-2}$ Hz). 
There are
no sharp features or discontinuities in \epft in the
range $\sim 64-1000$ s ($1.6 \times 10^{-2}$ to $10^{-3}$ Hz ). 
Note that if variability ceased somewhere in the range
$T \sim 64-1000$ s
it would have been detected. Yet $EPF(T=64 \ s)$ is
$\sim 1.78 \sigma$ away from the value for a constant source so
such a sharp break can be ruled out at $>92\%$ confidence.

The data are in fact compatible with
$f \propto T^{\Gamma}$. Figure 4 (solid line) shows
a comparison of the data with a model with $\Gamma = 0.435$ 
(normalization $8.2 \times 10^{-3}$). 
These parameters were obtained
from an `eyeball fit' and no attempt was made at formal fitting. 
Note that $\Gamma$ is related to the power-law index of the
power spectrum, $\alpha$, by $\Gamma = -(1 + \alpha)$. This is
because \epft is a function of the number of variable bins
picked at a given frequency interval $1/T$, and thus is a function of 
the power in that frequency interval. The power spectrum is defined
as power per unit frequency so the power-law index for
fitting \epft must be lowered by unity.  Thus the
data are consistent with $\alpha = -1.435$ in the
range $5 \times 10^{-2}$ to $10^{-3}$ Hz, which is compatible with 
\exosat measurements of MCG $-$6$-$30$-$15 at lower frequencies 
(e.g. $\alpha = -1.48^{+0.19}_{-0.18}$ by Lawrence and Papadakis 1993
and $\alpha = -1.36^{+0.25}_{-0.32}$ by Green \etal 1993).

If the variability that is observed down to $T=48$ s ($2.1 \times 10^{-2}$
Hz) is due to modulation by matter orbiting the putative black hole,
then we can infer an upper limit on the black-hole mass by requiring
that the emitting matter is not located closer than 
the radius for marginally
stable circular orbits (i.e. 3 Schwarzschild radii). This
condition means that the observed period, $P$, 
for Keplerian 
orbits around a Schwarzschild and a maximally rotating Kerr black hole
is $P \ge 1.05 \times 10^{4} M_{8} $ s and $P \ge 4.6 \times 10^{4} M_{8}$ s
respectively 
($M_{8}$ is the mass in units of $10^{8} M_{\odot}$). 
Then $P=48$ s implies  $M_{8} < 4.6 \times 10^{-3}$
for a Schwarzschild metric and $M_{8} < 10^{-3}$  for a maximally
rotating Kerr metric. These masses are rather small and imply
luminosities in excess of the Eddington limit.
This argues against the observed variability being due to orbital
modulation. Note that the upper limits on the black hole masses 
obtained by Green \etal (1993) are in error by a factor of 10 because
the numerical factor in their equation 6 is incorrect and should be
$10^{3}$ and not $10^{4}$. Also, they arbitrarily assume an orbital
radius of 5 Schwarzschild radii. These factors, along with the fact that
the \asca data detect variability at shorter timescales than the
\exosat data, explain why our upper limit for the black-hole mass
in MCG -6-30-15 (under the assumption of orbital modulation around
a Schwarzschild hole) is a factor of 30 less than the Green \etal
(1993) estimate of $1.4 \times 10^{7} M_{\odot}$.

Clearly, to infer a source size from the fastest observed variability
is too simplistic. 
Instead, we have begun, for the first time, to
probe the source sub-structure since
the rapid variability may be produced by localized
flares or hotspots.
The lower limit on any break frequency of
$1.6 \times 10^{-2}$ Hz (64 s) represents an upper limit
($1.92 \times 10^{12} \ \rm cm$) only to the size
of a region responsible for the shortest observed (and low-amplitude)
variability, and not necessarily to the size of the source region
as a whole. 

\section{CONCLUSIONS}
\label{conc} 

We have presented a new method for analysing gapped time-series 
which is capable of probing the power-spectrum of AGN up to
$\sim 5 \times 10^{-2}$ Hz for sources with count rates of the order $\sim 1$
ct/s or more. We have applied the method to a four-day \asca observation
of the Seyfert 1 galaxy \mcg  to search for a high-frequency cut-off
in the power spectrum. The data are consistent with a 
power-law power spectrum, with no break 
in the range $10^{-3}$ Hz to $1.6 \times 10^{-2}$ Hz (at a
confidence level of 92\%). The power-law index 
of the power spectrum in this range is consistent with $\sim -1.4$.
In the range $1.6 \times 10^{-2}$ Hz to $5 \times 10^{-2}$ Hz,
the fluctuations in the intrinsic mean count rate of the 
source can be constrained to be less than 7\% at greater
than 99\% confidence. It is likely that the observed variability
is {\it not} due to orbital modulation since black hole masses
of less than $5 \times 10^{5} M_{\odot}$ and super-Eddington
luminosities are implied.

\vspace{2cm}

The authors would like to thank Karen Leighly for some extremely
useful discussions and an anonymous referee for some very helpful
criticism.

\section*{Figure Captions}

\noindent
{\bf Figure 1} \\
The `Excess Pair Fraction', $EPF(\mu)$, calculated from 
simulations
using $10^{6}$ bin-pairs. The mean number of counts in each bin is
$\mu(1+ f \eta)$, where $\eta$ is a uniform
random deviate between $-1$ and $+1$ and
$f$ represents a fractional variability parameter. 
The curves are shown for several
values of $f$ in the range 0.0--1.0 (a value of 0.0 corresponds to
a constant source). 
The error bars are obtained from the square root of the number
of bin-pairs contributing to $EPF(\mu)$,
divided by the total number of bin-pairs.
Note that although the error bars are plotted,
they are too small to be visible for most of the parameter space.
See text for details.

\noindent
{\bf Figure 2} \\
The `Excess Pair Fraction', $EPF(T)$, calculated from
simulations of a source with mean count rate 1.0 ct/s. Good Time Intervals
(GTI) were taken from a real, $\sim 4.2$ day  \asca observation of MCG $-$6$-$30$-$15
(see Figure 3). Panel (a) shows the case for a constant source and
panel (b) shows the same source with a 3\% sinusoidal modulation
(6\% peak-to-peak) and a period of 367 s. 

\noindent
{\bf Figure 3} \\
The 0.5--10 keV SIS0 lightcurve for the extended \asca observation of
MCG $-$6$-$30$-$15. The lightcurve is binned at 512 s. The reference
time is 23 July 1997 UT 05:56:13.

\noindent
{\bf Figure 4} \\
The `Excess Pair Fraction' ($EPF$) calculated from the 
extended \asca observation of 
MCG $-$6$-$30$-$15 (filled circles). 
The data from the two SIS instruments (0.5--10 keV) and
two GIS instruments (0.7--10 keV) were used.
For bin sizes less than 32 s
only GIS data were used, the $EPF$ being averaged for GIS2 and GIS3.
Otherwise, the $EPF$ was averaged for all four \asca instruments.
The error bars are obtained from the square root of the number
of bin-pairs contributing to $EPF$, 
divided by the total number of bin-pairs tested.
The dotted lines show the predicted EPF for $f = 0\%$ (constant source),
7\% and 20\% (as indicated), where $f$ is the maximum bin-to-bin
fluctuation in the mean count rate of the source.
The mean number of counts in each bin of width $T$ is
$\mu_{0}T(1+ f \eta)$ where $\eta$ is a uniform
random deviate between $-1$ and $+1$. The solid curve
corresponds to a model with $f = 8.2 \times 10^{-3} T^{+0.435}$. See text 
for details. 


\newpage

\begin{references}

\reference{BaoO1994} Bao, G., \& Ostgaard, E. 1994, ApJ, 422, L51

\reference{Czer1997} Czerny, B., \& Lehto, H. J. 1997, MNRAS, 285, 365

\reference{Fior1989} Fiore, F., Massaro, E., \& Perola, G. C. 1989, ApJ, 347, 171

\reference{Gend1995} Gendreau, K. C., \etal 1995, PASJ, 47, L5

\reference{Gree1993} Green, A. R., McHardy, I. M., \& Lehto, H. J. 1993,
MNRAS, 265, 664 

\reference{Iwas1996} Iwasawa, K., \etal 1996, MNRAS, 282, 1038

\reference{Lawr1993} Lawrence, A., \& Papadakis, I. E. 1993, ApJ, 414, L85

\reference{McHa1987} McHardy, I. M., \& Czerny, B. 1987, Nat, 325, 696

\reference{Merr1994} Merrifield, M., \& McHardy, I. M. 1994, MNRAS, 271, 899 

\reference{Nand1997} Nandra, K., George, I. M., Mushotzky, R. F., Turner, T. J., \&
Yaqoob, T. 1997, ApJ, 476, 70 

\reference{Otan1996} Otani, C., \etal 1996, PASJ, 48, 211
 
\reference{Papa1993} Papadakis, I. E., \& Lawrence, A. 1993, Nat, 361, 233

\reference{Papa1995} Papadakis, I. E., \& Lawrence, A. 1995, MNRAS, 272, 161

\reference{Papm1995} Papadakis, I. E., \& McHardy, I. M. 1995, 273, 923 

\reference{Tana1994} Tanaka, Y., Inoue, H., \& Holt, S. S. 1994, PASJ, 46, L37

\reference{Tana1995} Tanaka, Y., \etal 1995, Nat, 375, 659

\end{references}
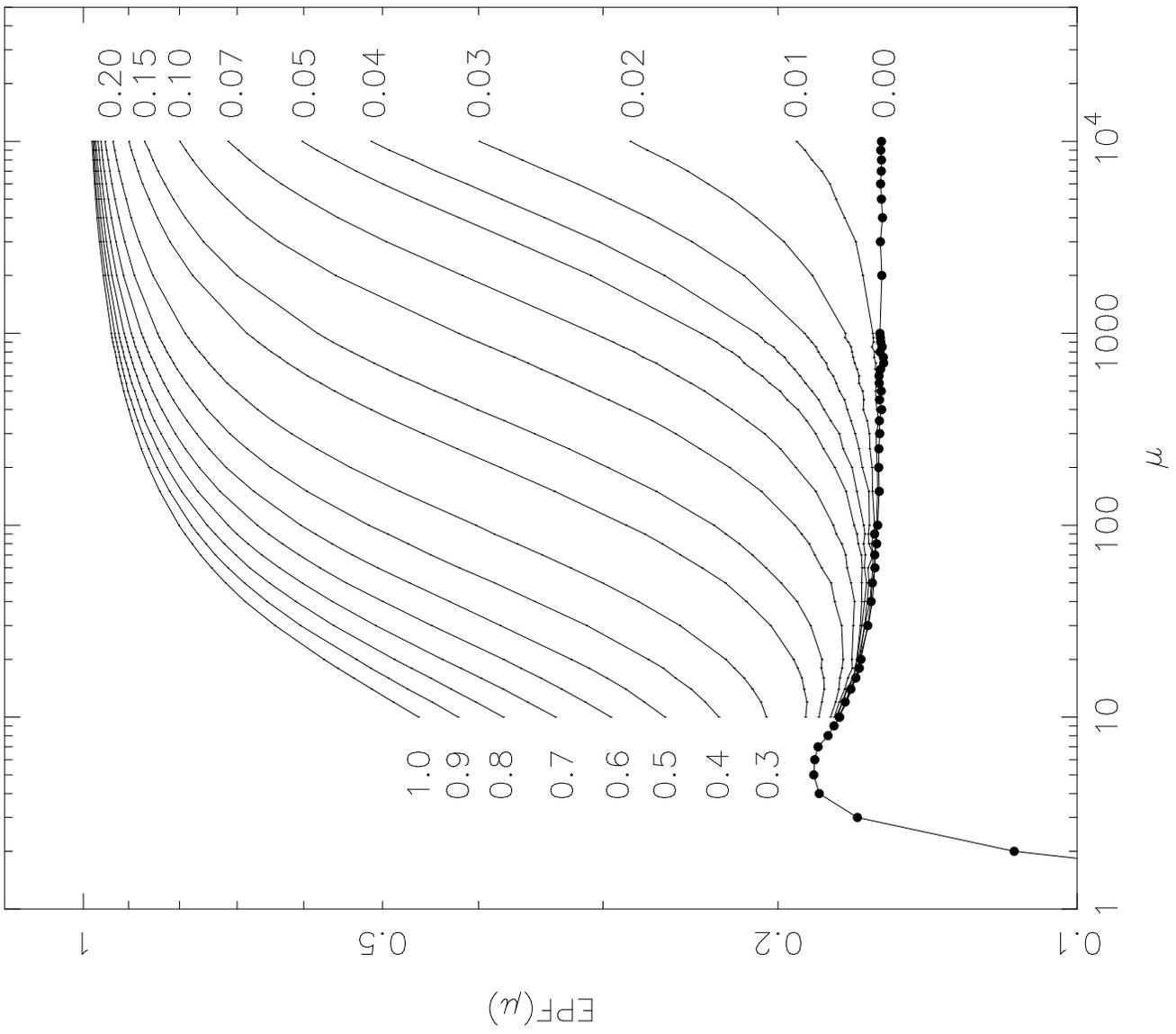
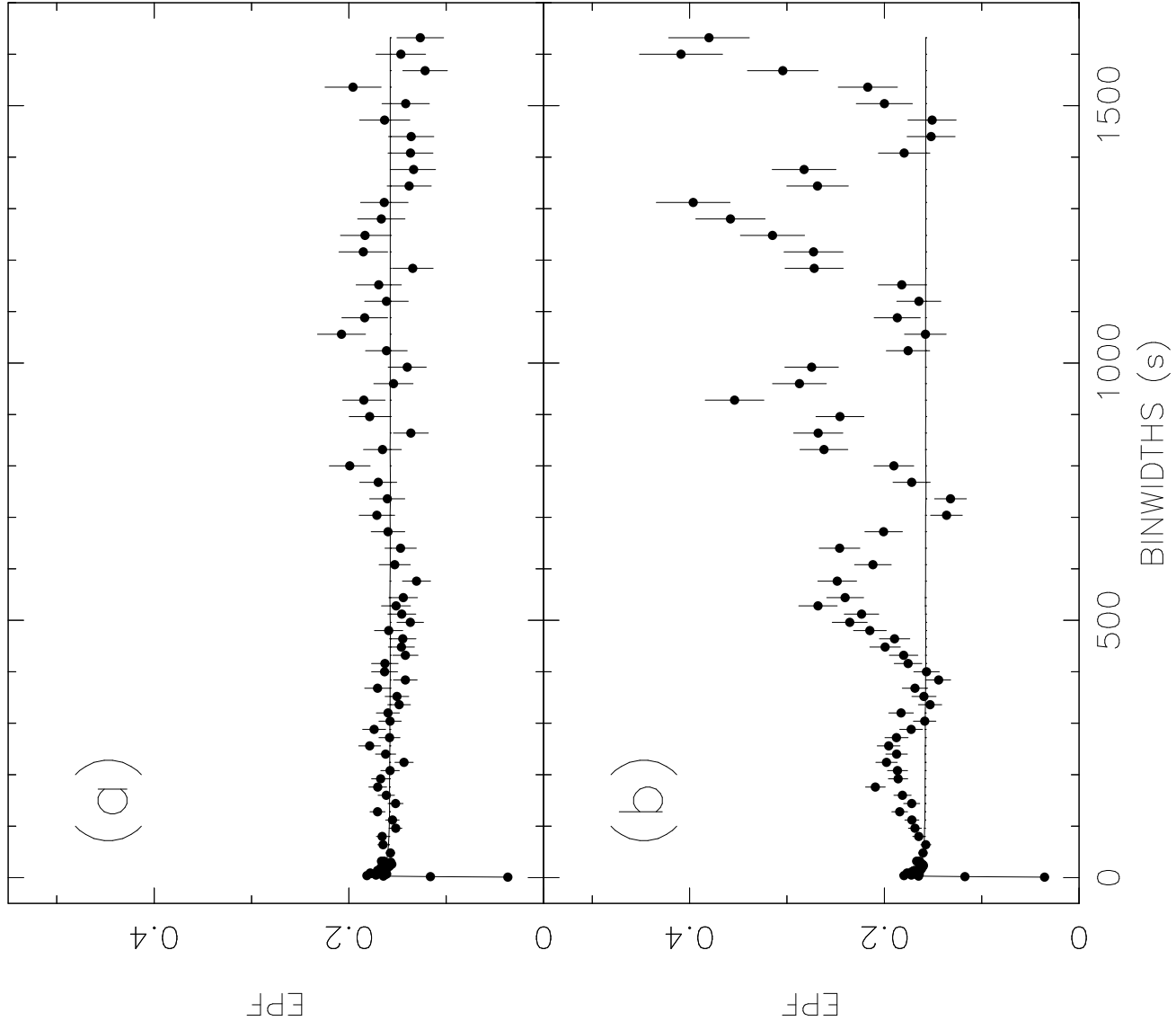
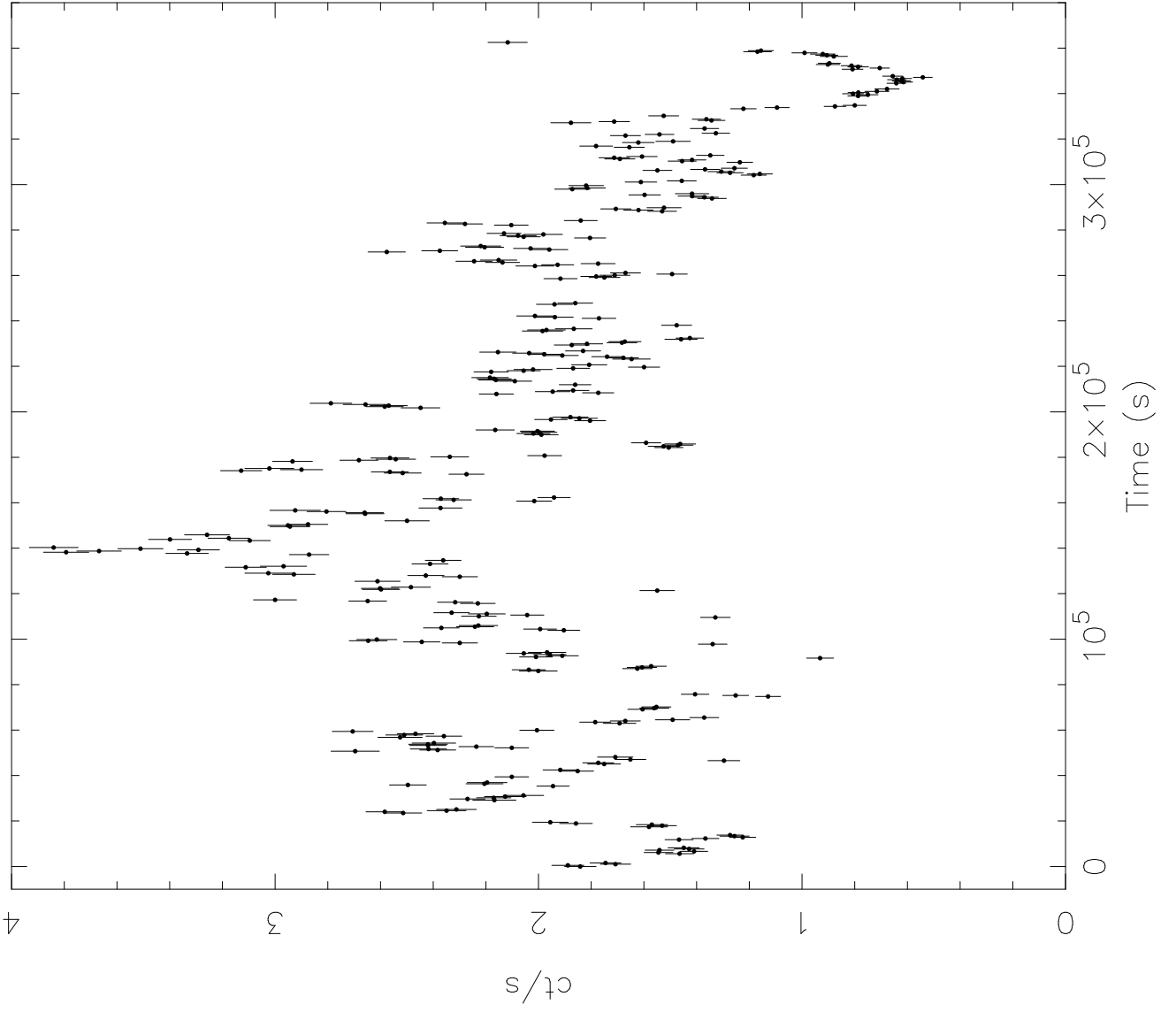
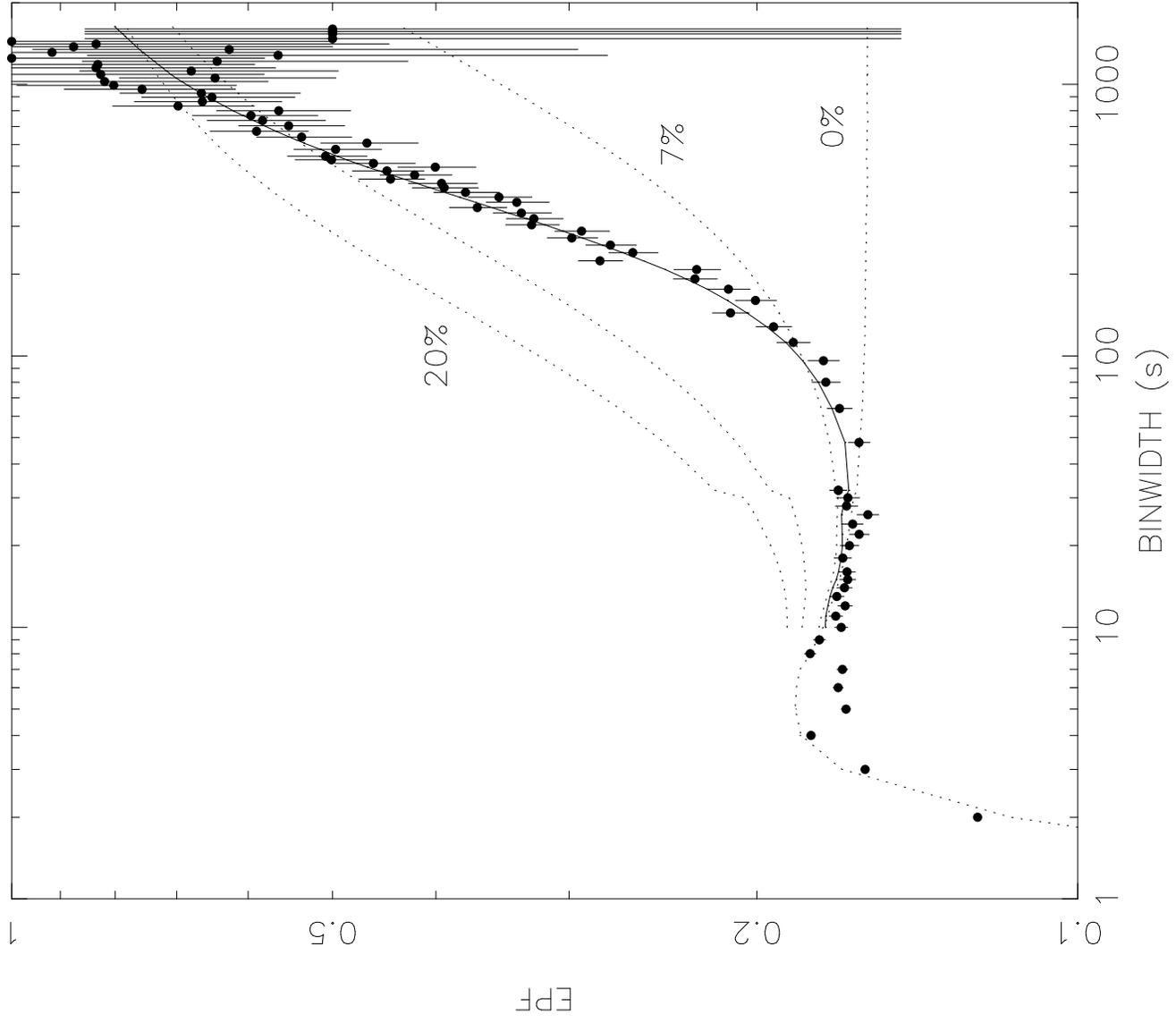
\end{document}